\documentclass[12pt,amstex,aps,amsfonts,prsty]{article}
\usepackage[dvips]{graphicx}
\usepackage{amssymb}
\usepackage{amsmath}
\usepackage{a4wide}
\usepackage{citesort}

\newcommand{\seq}[2][n]{\lbrace #2_{1},#2_{2},#2_{3},\ldots,\,#2_{#1} \rbrace}
\def\be{\begin{equation}}
\def\fin{\end{equation}}
\def\disp{\displaystyle}

\def\R{{\mathbb{R}}}

\def\T{{\sf T\kern-.45em T}}
\def\C{\kern.1em{\raise.47ex\hbox{$\scriptscriptstyle |$}}
             \kern-.40em{\sf C}}

\def\hfl{\disp\mathop{\hbox to 10mm{\rightarrowfill}}}

\newcommand{\euler}[2]{\left<\genfrac{}{}{0pt}{}{#1}{#2}\right>}

\begin{document}

\begin{titlepage}

\begin{center}

{\LARGE \bf Random walk generated by random permutations of
$\bf \{1,2,3, \ldots,n+1\}$.}

\vspace{0.4in}

{\Large \bf G.Oshanin$^{1,2}$ and R.Voituriez$^{1}$}

\vspace{0.3in}

{\sl $^{1}$ Laboratoire de Physique Th{\'e}orique des Liquides, \\
Universit{\'e} Paris 6, 4 Place Jussieu, 75252 Paris, France }

{\sl $^{2}$ Max-Planck-Institut f\"ur Metallforschung,
Heisenbergstr. 3, D-70569 Stuttgart, Germany, and Institut f\"ur
Theoretische und Angewandte Physik, Universit\"at Stuttgart,
Pfaffenwaldring 57, D-70569 Stuttgart, Germany}

\vspace{0.3in}

\begin{abstract}
We study properties of a non-Markovian random walk $X^{(n)}_l$, $l
=0,1,2, \ldots,n$, evolving in discrete time $l$ on a
one-dimensional lattice of integers, whose moves to the right or
to the left  are prescribed by the \text{rise-and-descent}
sequences characterizing random permutations $\pi$ of $[n+1] =
\{1,2,3, \ldots,n+1\}$. We determine exactly the probability of
finding the end-point $X_n = X^{(n)}_n$ of the trajectory  of such a
permutation-generated random walk (PGRW) at site $X$, and show
that in the limit $n \to \infty$ it converges to a normal distribution
with a smaller, compared to the conventional P\'olya random walk,
diffusion coefficient. We formulate, as well, an auxiliary
stochastic process whose distribution is identic to the
distribution of the intermediate points $X^{(n)}_l$, $l < n$, which
enables us to obtain the probability measure of different
excursions and to define the asymptotic distribution of the number
of "turns" of the PGRW trajectories.

\end{abstract}

\vspace{0.3in}

PACS: 02.50.-r Probability theory, stochastic processes, and
statistics;\\ 05.40.-a Fluctuation phenomena, random processes,
noise, and Brownian motion

\end{center}

\end{titlepage}

\pagebreak

\section{Introduction.}

Properties of random unrestricted or certain patterns-avoiding
permutations of  $[n] = \{1,2,3, \ldots, n\}$ have been analyzed
by mathematicians in group theory and combinatorics for many years
\cite{perm1,perm2,perm3}. Studies of several problems emerging
within this context, such as, e.g., the celebrated Ulam's longest
increasing subsequence problem (see, e.g., Refs.\cite{baik,wilf,od}
and references therein), provided an entry to a rich and diverse
circle of mathematical ideas \cite{dia}, and were also found
relevant to certain physical processes, including random surface
growth \cite{sep,pra,jon,nec} or 2D quantum gravity (see, e.g.,
Ref.\cite{od}).

In this paper we focus on random unrestricted\footnote{In the
sense that all patterns are permitted.} permutations from a bit
different viewpoint addressing the following question: what sort
of random walk one gets when random uniform permutations are used
as a generator of the walk? Here we consider a simple model of
such a permutation-generated random walk (PGRW), evolving in
discrete time on a one-dimensional lattice of integers, in which
model the moves of the walker to the right or to the left are
prescribed by the \text{rise-and-descent} sequence characterizing
each given permutation $\pi = \{\pi_1, \pi_2, \pi_3, \ldots , \pi_l, \ldots , \pi_{n+1}\}$. In a standard notation, the "rises"
(or the "descents") of the permutation $\pi$  are such
values of $l$ for which $\pi_l < \pi_{l+1}$ ($\pi_l > \pi_{l+1}$) \cite{1}.
We note that such a generator is evidently different of those
producing conventional RWs, since here a finite amount $n+1$ of
numbers is being shuffled and moreover, neither of any two numbers
in each permutation may be equal to each other; this incurs, of
course, correlations in the \text{rise-and-descent} sequences and
implies that the resulting PGRW is a non-Markovian process.

In this paper we determine exactly many characteristic properties
of such a random walk. First, we define the probability ${\cal
P}_n(X)$ of finding the end-point $X_n$ of the PGRW trajectory at
site $X$. We show next that in the long-time limit ${\cal P}_n(X)$
converges to a normal distribution in which the effective
diffusion coefficient $D = 1/6$ is three times smaller than the
diffusion coefficient ($D = 1/2$) of the conventional P\'olya
random walk in one dimension (1D), which signifies that
correlations in the generator are marginally important. Indeed,
assuming that correlations in the sequence of rises depend only on
the relative distance between their positions, we deduce from
${\cal P}_n(X)$ the two- and four-point correlation functions
explicitly and demonstrate that correlations extend
 effectively
to nearest-neighbors only. Next, at a closer look on the intrinsic
features of the PGRW trajectories, we formulate an auxiliary
Markovian stochastic process. We show that, despite the fact that
this process is Markovian, while the PGRW is not a Markovian
process, the distribution of the auxiliary stochastic process
appears to be identic to the distribution of the intermediate
states of the PGRW trajectories. This enables us to obtain the
probability measure of different excursions of the PGRW, deduce a
general expression for $k$-point correlation functions in the
\text{rise-and-descent} sequences, as well as to evaluate the
asymptotic form of the distribution of the number of "turns" of
the PGRW trajectories, i.e. the number of times the walker changes
the direction of its motion up to time $n$. Finally, we discuss,
in the diffusion limit, the continuous space and time version of
the PGRW.

The paper is outlined as follows: In Section 2 we formulate the
model more specifically. In Section 3 we derive the distribution
function of the end-points of the walker's trajectories exactly and
analyse its asymptotical behavior.
In Section 4 we consider the
correlations in the \text{rise-and-descent} sequences and obtain
explicit results for two- and four-point correlation
functions.
Further on, in Section 5 we introduce an auxiliary stochastic
process which has the same distribution as the intermediate points
of the walkers trajectory $X_l^{(n)}$, $l=1,2, \ldots, n-1$, and
obtain the probability measure of different excursions.
In Section 6 we determine the asymptotic form of the distribution function of the number of "turns" of the PGRW trajectory, relating it to $k$-point correlation functions in the \text{rise-and-descent} sequences.
Next, in Section
7 we discuss the continuous space and time analogue of the PGRW.
Finally, in Section 8 we conclude with a brief summary of our
results and discussion.

\section{Model.}

Let $\pi = \seq[n+1]{\pi}$ denote a random, unconstrained
permutation of
$[n+1]$. We
rewrite it next in two-line notation as
\begin{equation}
\begin{pmatrix}
1&2&3& \dots& n+1\\\pi_1&\pi_2&\pi_3& \ldots &\pi_{n+1}
\end{pmatrix}
,
\end{equation}
where the numbers in the first line will be thought of as the
values attained by a discrete "time" variable $l$. We call, in a
standard notation,  as the "rise" (or the "descent") of the
permutation $\pi$, such values of $l$ for which $\pi_l <
\pi_{l+1}$ ($\pi_l > \pi_{l+1}$) \cite{1}.

Consider now a walk evolving in time $l$ on an infinite in both
directions  one-dimensional lattice of integers according to the
following rules:

-at time moment $l = 0$ the walker is placed at the origin.

-at time moment $l = 1$  the walker is moved one step to the right
if $\pi_1 < \pi_2$, i.e., $l = 1$ is a rise, or to the left if
$\pi_1
> \pi_2$, i.e $l = 1$ is a descent.

-at time moment $l = 2$,  the walker is moved one step to the
right (left) if $\pi_2 < \pi_3$ ($\pi_2 > \pi_3$, resp.) and etc.

Repeated $l$ times, this procedure results in a random,
permutation-dependent trajectory $X_l^{(n)}$,
 $(l = 1,2, \ldots , n)$,
\begin{equation}
\label{def}
X_{l}^{(n)} = \sum_{k=1}^{l} s_k,
\end{equation}
where the "spin" variable $s_k$ is \be \label{spin} s_k =
\theta(\pi_{k+1}-\pi_{k}),
\end{equation}
$\theta(x)$ being the theta-function with the properties
\begin{equation}
\label{theta}
\theta(x) \equiv
\begin{cases}
+1& \text{if $x > 0$},\\
-1& \text{otherwise}.
\end{cases}
\end{equation}
The questions, which we address here, are a) the form of the
probability ${\cal P}_n(X)$ of finding such a walker at position
$X$ at time moment $n$, b) asymptotical behavior of ${\cal
P}_n(X)$, c) the probability measure of different excursions
$X_{l}^{(n)}$ and d) the distribution function of the number of
"turns" of the PGRW trajectories. As a by-product of our analysis,
we also determine some specific correlations in permutations,
embodied in the moments of the walker's trajectories., i.e. the
$k$-point correlation functions in the \text{rise-and-descent}
sequences, as well as the asymptotic distribution of the number
of peaks and throughs in random permutations.

\section{The probability distribution.}

One possible way to determine ${\cal P}_n(X)$ is to reconstruct it
from the moments $<X_{n}^{2 q}>$, ($q = 1,2,3, \dots$), of the
end-point $X_{n}$ of the walker's trajectory, $X_n = X_{l=n}^{(n)}
= \sum_{k=1}^{n} s_k,$
\begin{equation}
\label{mom} \left\langle X_{n}^{2 q} \right\rangle =
\frac{1}{(n+1)!} \sum_{\{\pi\}} \left[\sum_{l=1}^{n}
s_l \right]^{2q},
\end{equation}
where
the angle brackets denote "averaging" procedure - summation
with respect to the set
$\{\pi\}$
of all possible permutations $\pi$,
weighted by their total number  $(n+1)!$.

Pursuing this kind of approach, we may then represent the
summation over $\{\pi\}$
as a multiple
summation over
states of
a set of independent
variables $\{a_i\}$,
each running from $1$ to
$n+1$:
\begin{equation}
\sum_{\{\pi\}} \ldots = \lim_{J \to \infty}
\sum_{a_1 = 1}^{n+1} \sum_{a_2=1}^{n+1} \ldots
\sum_{a_{n+1} = 1}^{n+1} \exp\left( - J \sum_{(i,j)} \delta_{a_i,a_j} \right)  \ldots ,
\end{equation}
where the sum with the subscript
$(i,j)$
 extends over
all $i,j$ pairs such that $i,j \in 1,2,3, \ldots , n+1$,
excluding $i
= j$, $\delta_{a_i,a_j}$
is the Kroneker-symbol, while the factor
\begin{equation}
\lim_{J \to \infty} \exp\left( - J \sum_{(i,j)} \delta_{a_i,a_j} \right) =
\prod_{(i,j)} \left(1 -
\delta_{a_i,a_j}\right)
\end{equation}
accounts for the constraint that
$\pi_i \neq \pi_j$ for any $i$ and $j$, $i \neq j$.

Therefore, calculation of moments of $X_n$ amounts to the
evaluation of (rather) special correlations in an $(n+1)$-state
Potts-like model with long-range interactions on the 1D
 lattice containing $n+1$ sites,
which may be approached, in particular,
using a non-trivial transformation of
variables proposed in Ref.\cite{vik}.
It appears, however, that within this approach
the calculation of already the
second
moment $<X_{n}^{2}>$ is quite laborious, although the final result is simple.

On the other hand, ${\cal P}_n(X)$ can be obtained
straightforwardly, in an explicit form, following a different way
of reasonings. To do this, let us recall that according to the
definition of the PGRW, at the $l$-th step the walker makes a move
to the right (left) if $l$ is the rise (descent) of the random
permutation $\pi$. Now, let ${\cal N}_{\uparrow}$ (${\cal
N}_{\downarrow}$) be the number of "rises" ("descents") in a given
random permutation $\pi$. Then, for this given permutation $\pi$
the end-point $X_{n}$ of the walker's trajectory is just
\begin{equation}
X_{n} = {\cal N}_{\uparrow} - {\cal N}_{\downarrow},
\end{equation}
which can be rewritten, using an evident "conservation law"
${\cal N}_{\uparrow} + {\cal N}_{\downarrow} = n$
as
\begin{equation}
\label{k}
X_{n} = 2 {\cal N}_{\uparrow} - n.
\end{equation}
Therefore, for a given permutation $\pi$, the
 end-point $X_{n}$
of the walker's trajectory $X_{l}^{(n)}$ is entirely $\textit
fixed$ by the value of the number of $\textit rises$ in this permutation.

Now, a total number  of permutations $\pi$ of $[n+1]$ having
$\textit exactly$ ${\cal N}_{\uparrow}$ rises is given by the
so-called Eulerian number \cite{2}:
\begin{equation}
\label{en}
\left \langle
\begin{matrix}
n+1 \\ \displaystyle  {\cal N}_{\uparrow}
\end{matrix}
\right \rangle
=
\sum_{r=0}^{{\cal N}_{\uparrow} + 1} (-1)^r {n+2 \choose r}
({\cal N}_{\uparrow} + 1
- r)^{n+1},
\end{equation}
where ${a \choose b}$ denotes the binomial coefficient.
Consequently,
in virtue of Eq.(\ref{k}),
we have that the probability ${\cal P}_{n}(X)$ of
finding the walker
at position $X$ at
time $n$, $(-n \leq X \leq n)$,
is given by
\begin{equation}
\label{eu}
{\cal P}_n(X) = \frac{[1 + (-1)^{X + n}]}{2 (n + 1)!}
\left \langle
\begin{matrix}
n+1 \\ \displaystyle \frac{X + n}{2}
\end{matrix}
\right \rangle.
\end{equation}
Using the representation of the Eulerian number
in Eq.(\ref{en}), 
the result in Eq.(\ref{eu})
can be also written down in explicit form
as the following finite
series:
\begin{equation}
\label{euu}
{\cal P}_n(X) = \frac{[1 + (-1)^{X + n}]}{2 (n+1)!}
\sum_{r=0}^{(X + n  + 2)/2} (-1)^r {n+2 \choose r} \left( \frac{X + n + 2}{2}
- r\right)^{n+1}.
\end{equation}

We derive next several useful integral
representations of the distribution function ${\cal P}_n(X)$ and
of the corresponding lattice Green function \cite{weiss}, which
will allow for an easy access to the large-$n$ asymptotical
behavior of ${\cal P}_n(X)$.

Making use of the equality \cite{abr}:
\begin{equation}
\label{jj} \frac{1}{(n+1)!} \sum_{r=0}^{[(n + 2 + b)/2]} (-1)^r
{n+2 \choose r} \left(\frac{n + 2 + b}{2} - r\right)^{n+1} =
\frac{2}{\pi} \int^{\infty}_0 \left(\frac{\sin(k)}{k}\right)^{n+2}
\cos(b k)
 dk,
\end{equation}
we find that ${\cal P}_n(X)$ admits the following compact form:
\begin{equation}
\label{int} {\cal P}_n(X) = \frac{[1 + (-1)^{X + n}]}{\pi}
\int^{\infty}_0 \left(\frac{\sin(k)}{k}\right)^{n+2} \cos(X k) d
k.
\end{equation}
Note that the integral representation in Eq.(\ref{int}) is
different from the usually encountered forms since here the upper
terminal of integration is infinity. As a matter of fact, an
integral representation of ${\cal P}_n(X)$ in Eq.(\ref{eu}), in
 which the integration extends over the first Brillouin
zone \textit{only}, has a completely different form, compared to that
in Eq.(\ref{int}), and reads
\begin{eqnarray}
\label{br1}
&&{\cal P}_n(X) = \frac{(-1)^{n+1}}{(n+1)! \pi} \int_{0}^{\pi}
 \left( \sin^{n+2}(k) \frac{d^{n+1}}{d k^{n+1}} \cot(k) \right) \cos(X k) dk = \nonumber\\
&=&  \frac{1}{\pi^{n+3} (n+1)!} \int^{\pi}_0 \left[
\Psi_{n+1}\left(1-\frac{k}{\pi}\right) +
(-1)^{n} \Psi_{n+1}\left(\frac{k}{\pi}\right)\right] \sin^{n+2}(k) \cos(X k) dk,
\end{eqnarray}
where $\Psi_n(k)$ is the polygamma function \cite{abr}. Derivation
of the result in Eq.(\ref{br1}) is outlined in the Appendix A.

Consider finally the asymptotic forms of the probability
distribution function. To do this, it is expedient to turn to the
lattice Green function ${\cal G}(X,z)$ associated with the distribution function of the end-point of the PGRW trajectories.
From
Eqs.(\ref{int}) and (\ref{br1}) we find that the lattice Green
function of the end-point of the PGRW is given by
\begin{eqnarray}
\label{br2} {\cal G}(X,z) = \sum_{n=0}^{\infty} {\cal P}_n(X) z^n
&=& \frac{2}{\pi} \int^{\infty}_0 \left(\frac{\sin(k)}{k}\right)^2
\frac{\cos(X k) dk}{\displaystyle \left(1  - z^2
\frac{\sin^2(k)}{k^2}\right)} =\nonumber\\
&=& \frac{1}{\pi z} \int^{\pi}_0 \frac{\sin\left(z \sin(k)\right)
\cos(X k)}{\sin\left(k - z \sin(k)\right)} dk.
\end{eqnarray}
Now, in the limit $z \to 1^{-}$, (which corresponds to the large-$n$
behavior of the distribution function ${\cal P}_n(X)$), one finds
that the leading behavior of ${\cal G}(X,z)$ is as follows:
\begin{equation}
\label{ss} {\cal G}(X,z \to 1^{-}) \sim \sqrt{\frac{3}{2 (1 - z)}}
\exp\left(- \sqrt{6 (1 - z) } |X|\right),
\end{equation}
which implies that in the asymptotic limit $n \to \infty$, the
probability ${\cal P}_{n}(X)$ of finding the end-point of the
walker's trajectory at position $X$ converges to a \textit{normal}
distribution:
\begin{equation}
\label{normal} {\cal P}_{n}(X) \sim  \left(\frac{3}{2 \pi
n}\right)^{1/2} \exp\left(- \frac{3 X^2}{2 n}\right).
\end{equation}
Therefore, in the limit $n \to \infty$ the correlations in the
generator of the walk - random permutations of $[n+1]$,  appear to
be marginally important; that is, they do not result in
\textit{anomalous} diffusion, but merely affect the "diffusion
coefficient" making it three times smaller than the diffusion
coefficient of the standard 1D P\'olya walk. We will consider the
form of correlations in the next section.

\section{Correlations in the \text{rise-and-descent} sequence.}

We address now a somehow
"inverse" problem:
that is, given the distribution
${\cal P}_n(X)$,
we aim to determine two- and four-point
correlations in the \text{rise-and-descent} sequences.

We start with the analysis of two-point correlations. From
Eq.(\ref{mom}) we represent the second moment of the walker's
displacement as
\begin{equation}
\label{q}
\left\langle X_{n}^2 \right\rangle =  \left\langle \left({\cal N}_{\uparrow}
- {\cal N}_{\downarrow} \right)^2 \right\rangle  =  \left \langle \left[\sum_{l=1}^{n}
s_l \right]^{2} \right\rangle = \displaystyle
 n + 2 \sum_{j_{1} = 1}^{n - 1} \sum_{j_2 = j_1 + 1}^{n} {\cal C}_{j_1,j_2}^{(2)},
\end{equation}
where we denote as ${\cal C}_{j_1,j_2}^{(2)}$ the two-point correlation function of
the \text{rise-and-descent} sequence,
\begin{equation}
\label{2}
{\cal C}_{j_1,j_2}^{(2)} = \displaystyle \Big \langle s_{j_1}
s_{j_2} \Big \rangle.
\end{equation}
From our previous analysis, we know
 already that $\langle X_{n}^2 \rangle \sim n/3$ as $n \to \infty$, (see
Eq.(\ref{normal})).
Consequently, in this limit the sum on the rhs of Eq.(\ref{q}) is expected to behave as
\begin{equation}
\sum_{j_1 = 1}^{n - 1} \sum_{j_2 = j_1 + 1}^{n} {\cal C}_{j_1,j_2}^{(2)} \sim - \frac{n}{3},
\end{equation}
where the sign "-" signifies that the \textit{rise-and-descent} sequence is predominantly
\textit{anticorrelated} and the probability of having two neighboring rises (or two
descents) is lower than the probability of having a
rise neighboring to a descent.
The question now is how do
the correlations decay
with a relative distance $m = |j_2 - j_1|$?
We will answer this question here assuming
that  ${\cal C}_{j_1,j_2}^{(2)}$ is function of the distance
$|j_2 - j_1|$ only, i.e.
\begin{equation}
{\cal C}_{j_1,j_2}^{(2)} \equiv {\cal C}_{j_2,j_1}^{(2)} = {\cal C}^{(2)}(m = |j_2 - j_1|).
\end{equation}
We note that such an assumption seems quite plausible at the first glance,
since the rises (and descents)
are evidently uniformly distributed on the interval $[1, n]$. We
will show in the next section
that it is actually the case  using an auxiliary stochastic process having the same
distribution of the intermediate steps as the PGRW.

Introducing the generating function of the form
\begin{equation}
{\cal X}^{(2)}(z) = \sum_{n=0}^{\infty} \left\langle X_{n}^2 \right\rangle z^{n},
\end{equation}
we get from Eq.(\ref{q})
  the following relation:
\begin{equation}
{\cal X}^{(2)}(z) = \frac{z}{(1 - z)^2} + \frac{2 z}{(1-z)^2}
{\cal C}^{(2)}(z),
\end{equation}
where ${\cal C}^{(2)}(z)$ is the generating function of two-point
correlations:
\begin{equation}
{\cal C}^{(2)}(z) = \sum_{m=1}^{\infty} z^{m} {\cal C}^{(2)}(m).
\end{equation}
Consequently, in order to define  ${\cal C}^{(2)}(m)$, we have to evaluate ${\cal X}^{(2)}(z)$.
To do this, we proceed as follows:
define the
Fourier-transformed ${\cal P}_n(X)$ as
\begin{eqnarray}
\tilde{{\cal P}}_n(k) = \sum_{X = - n}^{n} \exp\left(i k X\right)
{\cal P}_n(X).
\end{eqnarray}
Noticing first that ${\cal P}_n(n+1) \equiv 0$, then,
making use of the representation in Eq.(\ref{eu}) and changing the summation variable,
we get
\begin{equation}
\label{eu3} \tilde{{\cal P}}_n(k) = \frac{\displaystyle \exp\left(
- i k n\right)}{(n + 1)!} \sum_{r = 0}^{n+1} \exp\left(2 i k r
\right) \left \langle
\begin{matrix}
n+1 \\ \displaystyle r
\end{matrix}
\right \rangle.
\end{equation}
Further on, using the equality
\begin{equation}
\label{eu4}
\sum_{r = 0}^{n+1} y^{n + 1 - r} \left \langle
\begin{matrix}
n+1 \\ \displaystyle r
\end{matrix}
\right \rangle
= \left(1 - y\right)^{n+2} {\rm Li}_{-n-1}(y),
\end{equation}
where ${\rm Li}_{-n-1}(y)$ denotes the polylogarithm function
\cite{abr}, we obtain, by setting $y = \exp( - 2 i k)$,
\begin{equation}
\label{dd} \tilde{{\cal P}}_n(k) = \frac{(2 i)^{n+2}
\sin^{n+2}(k)}{(n + 1)!} {\rm Li}_{-n-1}\Big(\exp(- 2 i k)\Big).
\end{equation}
Next, taking advantage of the expansion \cite{abr}
\begin{equation}
{\rm Li}_{-n-1}(y) = (n + 1)! \left(\ln\frac{1}{y}\right)^{- n - 2} - \sum_{j=0}^{\infty} \frac{B_{n + 2 + j}}{j! (n + 2 + j)} \left(\ln y\right)^j,
\end{equation}
where $B_{j}$ stand for the Bernoulli numbers,
we arrive at the following result:
\begin{equation}
\label{ll} \tilde{{\cal P}}_n(k) =
\left(\frac{\sin(k)}{k}\right)^{n+2} \; \left[1 -
\begin{cases}
j_1(k)& \text{for $n$ odd},\\
j_2(k)& \text{for $n$ even},
\end{cases}
\right]
\end{equation}
with
\begin{equation}
\label{lll} j_1(k) = \frac{(-1)^{(n+1)/2} (2 k)^{n+3}}{(n+1)!}
\sum_{j=0}^{\infty} \frac{(-1)^j B_{n + 3 + 2j} }{(2 j + 1)! (n +
3 + 2j)} (2 k)^{2 j},
\end{equation}
and
\begin{equation}
\label{llll} j_2(k) = - \frac{(-1)^{n/2} (2 k)^{n+2}}{(n+1)!}
\sum_{j=0}^{\infty} \frac{(-1)^j B_{n + 2 + 2j} }{(2 j)! (n + 2 +
2j)} (2 k)^{2 j}.
\end{equation}

Now,  we aim to determine $\Big \langle X_n^2 \Big \rangle$ and
$\Big \langle X_n^4 \Big \rangle$ explicitly. Expanding
$\tilde{{\cal P}}_n(k)$ defined by Eqs.(\ref{ll}), (\ref{lll}) and
(\ref{llll}) into the Taylor series in powers of $k$, we find that
$\tilde{{\cal P}}_n(k)$ obeys
\begin{eqnarray}
\tilde{{\cal P}}_n(k) &=& 1 - \frac{1}{6} \Big(n + 2 - 2 \delta_{n,0}\Big) k^2 + \nonumber\\
&+&\frac{1}{360} \Big((5 n + 8) ( n + 2) -16 \delta_{n,0} - 24
\delta_{n,1} + 8 \delta_{n,2} \Big) k^4 + {\cal
O}\left(k^6\right),
\end{eqnarray}
which yields
\begin{equation}
\label{x2}
\Big \langle X_n^2 \Big \rangle \equiv \frac{1}{3} \Big(n + 2 - 2 \delta_{n,0}\Big)
\end{equation}
and
\begin{equation}
\label{x4}
\Big \langle X_n^4 \Big \rangle \equiv \frac{1}{15}
\Big((5 n + 8) ( n + 2) -16 \delta_{n,0} - 24 \delta_{n,1} + 8 \delta_{n,2} \Big).
\end{equation}
We note parenthetically that the obtained distribution ${\cal
P}_n(X)$ in Eq.(\ref{eu}) appears to be platykurtic, since the
kurtosis excess
\begin{equation}
\gamma = \frac{\Big \langle X_n^4 \Big \rangle}{\Big \langle X_n^2
\Big \rangle^2} - 3 = - \frac{6}{5 n} + {\cal
O}\left(\frac{1}{n^2}\right)
\end{equation}
is negative.

Now, from Eq.(\ref{x2}) we find
\begin{equation}
{\cal X}^{(2)}(z) = \frac{z \left(3 - 2 z\right)}{3 \left(1 - z\right)^2},
\end{equation}
which implies that ${\cal C}^{(2)}(z) = - z/3$ and hence, that
\begin{equation}
\label{jj}
{\cal C}^{(2)}(m) \equiv
\begin{cases}
\displaystyle -\frac{1}{3}& \text{if $m = 1$},\\
0& \text{if $m \geq 2$}.
\end{cases}
\end{equation}
Note that Eq.(\ref{jj}) signifies that for $m \geq 2$
the
two-point correlations ${\cal C}^{(2)}(m = |j_2 - j_1|)$ in the \text{rise-and-descent} sequence
decouple into the product $<s_{j_1}> <s_{j_2}>$ and hence,
vanish. Consequently, two-point correlations are short-ranged
 extending to  the nearest neighbors only.
Therefore, the probability $p_{\uparrow, \uparrow}(m)$
of having two rises (or two descents) a distance $m$
apart of each other is
\begin{equation}
\label{o1}
p_{\uparrow, \uparrow}(m) = p_{\downarrow, \downarrow}(m)
\equiv \frac{1 + {\cal C}^{(2)}(m)}{4} =
\begin{cases}
\displaystyle \frac{1}{6}& \text{if $m = 1$},\\
\displaystyle \frac{1}{4}& \text{if $m \geq 2$},
\end{cases}
\end{equation}
while
\begin{equation}
\label{o2}
p_{\uparrow, \downarrow}(m) = p_{\downarrow, \uparrow}(m)
\equiv \frac{1 - {\cal C}^{(2)}(m)}{4} =
\begin{cases}
\displaystyle \frac{1}{3}& \text{if $m = 1$},\\
\displaystyle \frac{1}{4}& \text{if $m \geq 2$}.
\end{cases}
\end{equation}

Consider next behavior of the four-point correlations:
\begin{equation}
\label{4}
{\cal C}^{(4)}_{j_1,j_2,j_3,j_4} = \Big \langle s_{j_1} s_{j_2}
s_{j_3} s_{j_4}\Big \rangle,
\end{equation}
where $j_1 < j_2 < j_3 < j_4$. From our result in Eq.(\ref{jj}) it seems natural to expect that
${\cal C}^{(4)}_{j_1,j_2,j_3,j_4}$ vanishes as soon either (or both) $j_2 - j_1$ or $j_4 - j_3$ are greater than unity. Consequently,
only non-vanishing four-point correlations
are of the form
\begin{equation}
{\cal C}^{(4)}(m) = \Big \langle s_{j_1} s_{j_1+1}
s_{j_1+m+1} s_{j_1+m+2}\Big \rangle.
\end{equation}
To evaluate ${\cal C}^{(4)}(m)$, we will proceed along
the same lines as we
did with the analysis of two-point correlations.
From Eq.(\ref{mom}) we have
\begin{eqnarray}
\label{ju}
\left\langle X_{n}^4 \right\rangle &=&  \left\langle \left({\cal N}_{\uparrow}
- {\cal N}_{\downarrow} \right)^4 \right\rangle  =  \left \langle \left[\sum_{l=1}^{n}
s_{l}\right]^{4} \right\rangle = \nonumber\\
&=& 4! \sum \frac{\displaystyle \Big< s^{m_1}_1
s^{m_2}_2 \cdots s^{m_n}_n\Big>}{\displaystyle m_1! m_2! \cdots m_n!},
\end{eqnarray}
where the sum extends over all positive integer
solutions of equation $m_1 + m_2 + \cdots + m_n = 4$.
Taking into account our results for ${\cal C}^{(2)}(m)$,
as well as noticing that  $s^4_{l} = s^2_{l} \equiv 1$, while
$s^3_{l} = s_{l}$ for any $l$,
we find that
\begin{eqnarray}
&&\left\langle X_{n}^4 \right\rangle = \frac{4!}{4!} n + \frac{4!}{2! 2!} \frac{n (n - 1)}{2} \theta(n-2) +
2 \frac{ 4!}{3! 1!} (n - 1) {\cal C}^{(2)}(1) \theta(n - 2) +  \nonumber\\
&+&  \frac{4!}{2! (1!)^2} \left(n (n - 3) + 2 \right) {\cal C}^{(2)}(1) \theta(n - 3) + 4! \sum_{m = 1}^{n - 3} \Big(n - 2 - m\Big)
{\cal C}^{(4)}(m) \theta(n - 4).
\end{eqnarray}
Multiplying both sides of the last equation by $z^n$ and performing summation, we find the following
relation between the generating function ${\cal X}^{(4)}(z)$ of the fourth moment of the walker's displacement
and the generating
function ${\cal C}^{(4)}(z)$ of the four-point correlations:
\begin{equation}
{\cal X}^{(4)}(z) = \frac{15 z  + 35 z^2 - 80 z^3}{15 (1 - z)^3} + \frac{4! z^3}{(1 - z)^2} {\cal C}^{(4)}(z).
\end{equation}
On the other hand, from Eq.(\ref{x4}) we have that
\begin{equation}
{\cal X}^{(4)}(z) = \frac{z \left(35 z  + 15 - 80 z^2 + 48 z^3 - 8 z^4\right)}{15 (1 - z)^3},
\end{equation}
which yields, eventually,
\begin{equation}
{\cal C}^{(4)}(z) = \frac{z (6 - z)}{45 (1 - z)} = \frac{2}{15} z +
\frac{1}{9} \left(z^2 + z^3 + z^4 + \cdots\right),
\end{equation}
and hence,
\begin{equation}
\label{zzz}
{\cal C}^{(4)}(m) =
\begin{cases}
\displaystyle \frac{2}{15}& \text{if $m = 1$},\\
\displaystyle \frac{1}{9}& \text{if $m \geq 2$}.
\end{cases}
\end{equation}
Note that Eq.(\ref{zzz})
signifies that the four-point correlations decouple into the product of nearest-neighbor
two-point correlations, ${\cal C}^{(4)}(m) =
{\cal C}^{(2)}(1) {\cal C}^{(2)}(1)$, for $m \geq 2$.

Finally, noticing that
three-point correlations of the
form $\Big\langle s_{j} s_{j+1}
s_{j+2}$
are equal to zero, we may straightforwardly calculate
the probabilities of several particular configurations involving three and four rises and descents.
In particular, the probability of having two neighboring rises
and a descent at distance $m$ apart of them is
given by
\begin{eqnarray}
\label{p0}
p_{\uparrow \uparrow, \downarrow}(m) &=& \frac{1}{8}
\Big\langle \left(1+ s_{j} \right)
\left(1+ s_{j+1} \right)
 \left(1 - s_{j+m+1} \right) \Big\rangle = \nonumber\\
&=&
\begin{cases}
\displaystyle \frac{1}{8}& \text{if $m = 1$},\\
\displaystyle \frac{1}{12}& \text{if $m \geq 2$},
\end{cases}
\end{eqnarray}
while the probability of having two neighboring rises and another rise
at distance $m$ apart of them obeys:
\begin{eqnarray}
\label{p6}
p_{\uparrow \uparrow, \uparrow}(m) &=& \frac{1}{8}
\Big\langle \left(1+ s_{j} \right) \left(1+ s_{j+1} \right)
 \left(1 + s_{j+m+1} \right) \Big\rangle = \nonumber\\
&=&
\begin{cases}
\displaystyle \frac{1}{24}& \text{if $m = 1$},\\
\displaystyle \frac{1}{12}& \text{if $m \geq 2$}.
\end{cases}
\end{eqnarray}
In a similar fashion, we get that the probability of
having two rises and another pair of rises (descents) at
distance $m$ apart of them is given by
\begin{equation}
\label{p1}
p_{\uparrow \uparrow, \uparrow \uparrow}(m) =
\begin{cases}
\displaystyle \frac{1}{120}& \text{if $m = 1$},\\
\displaystyle \frac{1}{36}& \text{if $m \geq 2$},
\end{cases}
\end{equation}
the corresponding probabilities
involving three rises obey
\begin{equation}
\label{p2}
p_{\uparrow \uparrow,  \downarrow \uparrow}(m) = p_{\uparrow \downarrow,  \uparrow \uparrow}(m) =
\begin{cases}
\displaystyle \frac{3}{40}& \text{if $m = 1$},\\
\displaystyle \frac{1}{18}& \text{if $m \geq 2$},
\end{cases}
\;\;\; p_{\uparrow \uparrow,  \uparrow \downarrow}(m) = p_{\downarrow \uparrow,  \uparrow \uparrow}(m) =
\begin{cases}
\displaystyle \frac{1}{30}& \text{if $m = 1$},\\
\displaystyle \frac{1}{18}& \text{if $m \geq 2$},
\end{cases}
\end{equation}
while the probabilities involving two rises and two descents follow
\begin{equation}
\label{p3}
p_{\uparrow \downarrow,  \uparrow \downarrow}(m) =
\begin{cases}
\displaystyle \frac{2}{15}& \text{if $m = 1$},\\
\displaystyle \frac{1}{9}& \text{if $m \geq 2$},
\end{cases}
\;\;\; p_{\uparrow \uparrow, \downarrow \downarrow}(m) =
\begin{cases}
\displaystyle \frac{1}{20}& \text{if $m = 1$},\\
\displaystyle \frac{1}{36}& \text{if $m \geq 2$},
\end{cases}
\;\;\; p_{\downarrow \uparrow, \uparrow \downarrow}(m) =
\begin{cases}
\displaystyle \frac{11}{120}& \text{if $m = 1$},\\
\displaystyle \frac{1}{9}& \text{if $m \geq 2$}.
\end{cases}
\end{equation}
Equations (\ref{p0}), (\ref{p6}) to (\ref{p3}) show explicitly
that the probabilities of different configurations of the
\text{rise-and-descent} sequences depend not only on the number of
rises, but also on their order within the sequence.

\section{Trajectories ${\bf X^{(n)}_l}$ for ${\bf l < n}$.}

So far, we have defined the distribution of the end-points of the
trajectories $X^{(n)}_l$, but, apart from the explicit results on
the form of two- and four-point correlations, which allow us to
reconstruct trajectories $X^{(n)}_l$ with $n = 4$ (see Fig.1), we
do not have an access to information on the distribution of the
intermediate points $l = 1,2,3, \ldots, n-1$. On the other hand,
it seems to be quite non-trivial. Indeed, we deal with a random
walk, which makes a move of unit length at each moment of time
with probability $1$, but nonetheless looses somehow two thirds of
the diffusion coefficient. Given that the correlations are
short-ranged and extend to nearest-neighbors only, one might
expect that the PGRW behaves effectively as an
"\textit{antipersistent}" RW \cite{weiss} with a short-range
one-step memory, such that the probability, in view of the results
in Eq.(\ref{o1}) and (\ref{o2}), of making
on the $l$th step a move in the same
direction at which it
made a move on the $(l-1)$st step
is $1/3$, while
the probability of changing the direction is $2/3$. One readily
verifies that for such a walk $\Big< X_n^2 \Big> \sim n/2$, i.e.
the reduction in the "diffusion coefficient" is smaller than the
one we actually find for the PGRW, $(\Big< X_n^2 \Big> \sim n/3)$.
As a matter of fact, as we
observe in Fig.1, for the PGRW
the memory of the
"\textit{antipersistency}" is stronger and depends not only on the
number of steps to the right or to the left, which the walker has
already made, but also on their order. In other words, the PGRW
represents a genuine non-Markovian process.

\subsection{The probability
distribution of ${\bf X^{(n)}_l}$ for ${\bf l < n}$.}

To determine the structure of excursions $X^{(n)}_l$ of the PGRW
we adapt a method proposed by Hammersley \cite{hammersley} in his
analysis of the evolution of the longest increasing subsequence,
and elucidated subsequently in Ref.\cite{diaconis}. The basic idea
behind this approach, which we exploit here, is to build
recursively an auxiliary \textit{Markovian} stochastic process
$Y_l$, which is distributed exactly as $X^{(n)}_l$.

At each time step $l$, let us define a real valued random variable $x_{l+l}$,
uniformly distributed in $[0,1]$. Consider next a random walk on an
infinite in both directions
one-dimensional lattice of integers whose trajectory $Y_l$ is
constructed according to the following step-by-step process: at
each time moment $l$ a point-like particle is created at position
$x_{l+1}$. If $x_{l+1} > x_{l}$, a walker is moved one step to the
right; otherwise, it is moved one step to the left. The trajectory
$Y_l$ is then given by \be \label{Y} \disp
Y_l=\sum_{k=1}^{l} \theta(x_{k+1}-x_k),
\end{equation}
where $\theta(x)$ is the
theta-function defined in Eq.(\ref{theta}).
We note that the
joint process $(x_{l+1},Y_l)$, and therefore $Y_l$, are \textit{Markovian} since
 they depend only on $(x_{l},Y_{l-1})$. Note also that $Y_l$ is the sum of
{\it correlated} random variables; hence, one has to be cautious
when applying central limit theorems. A central limit theorem
indeed holds for the Markovian process  $Y_l$, but the summation rule
for the variance is not valid.

\begin{figure}[ht]
\begin{center}
\includegraphics*[scale=0.7, angle=0]{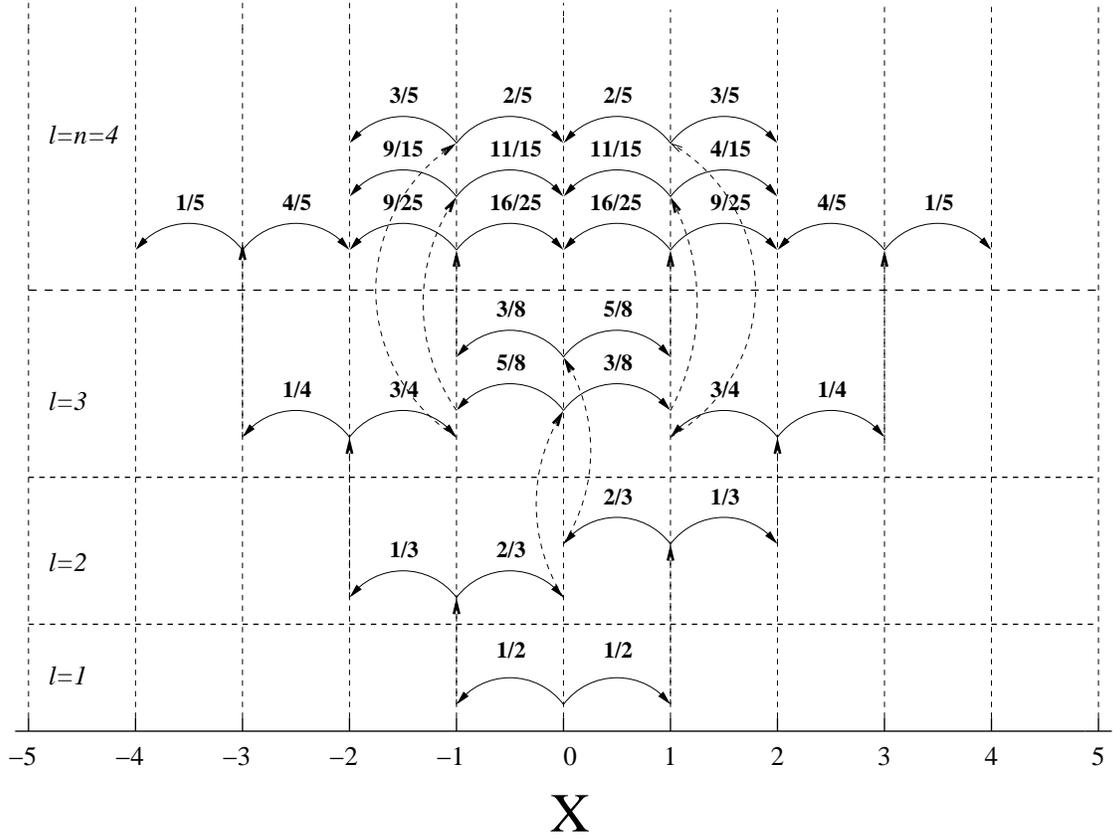}
\caption{\label{Fig2} {A set of all possible PGRW trajectories
$X_l^{(n)}$ for $n = 4$. Numbers above the solid arcs with arrows
indicate the corresponding transition probabilities. Dashed-lines with arrows connect the trajectories for
different values of $l$.}}
\end{center}
\end{figure}

Now, we aim to prove that  the probability ${\cal P}(Y_l = Y)$
that at time moment $l$ the trajectory $Y_l$ of the auxiliary
Markovian process appears on the site $Y$ has exactly the same
form as the Eulerian distribution obtained for the end-point of
the trajectory $X_{l}^{(n)}$. That is, ${\cal P}(Y_l = Y)$ obeys
\begin{equation}
\label{ue} {\cal P}(Y_l = Y) = \frac{[1 + (-1)^{Y + l}]}{2 (l +
1)!} \left \langle
\begin{matrix}
l+1 \\ \displaystyle \frac{Y + l}{2}
\end{matrix}
\right \rangle.
\end{equation}
To prove this statement, we assign a random permutation $\pi$ of
$[n+1]$ to each realization of sequence $\{x_l\}$, $l = 1, 2,
\ldots ,n + 1$, by ordering the $x_l$-s and requiring that $\pi_l$
is the row of $x_l$:
\begin{equation}
x_{\pi_1}<x_{\pi_2}<..<x_{\pi_{n+1}}
\end{equation}
Next, we note that according to the definition in Eq.(\ref{Y})
each random trajectory $Y_l$, $l \leq n$, is entirely determined
by ordering of the corresponding sequence $\{x_i\}_{i\le l+1}$.
This implies that $Y_l$ is adequately determined by the
permutation $\pi$ defined by such an  ordering procedure.

One writes next the weight $p(Y_l)$ of a given trajectory $Y_l$ as
an integral over the sequences $\{x_i\}_{i\le l+1}$ generating
$Y_l$, which can be represented as a sum over all corresponding
permutations: \be \disp p(Y_l)=\int_{\{x_i\}\ {\rm generates}\
Y_l} dx_1 \dots dx_{n+1} = \sum_{\pi \ {\rm generates}\ Y_l}
p(\pi), \fin where $p(\pi)$ is the probability of a given random
permutation $\pi$. On the other hand, $p(\pi)$ obeys:
\be\label{measure} \disp p(\pi)=\mathop{\int dx_1 \ldots
dx_{n+1}}_{x_{\pi^{-1}[\pi(k)-1]}\le x_k\le
x_{\pi^{-1}[\pi(k)+1]}}=\int_{0}^{1}dx_{n+1}\int_{0}^{x_{n+1}}dx_n..\int_{0}^{x_2}dx_1=\frac{1}{(n+1)!}
\end{equation}
where
$x_{\pi^{-1}(0)}=0$, $x_{\pi^{-1}(n+2)}=1$ and $\pi^{-1}$ denotes
the functional inverse of $\pi$, such that $\pi^{-1}(\pi_k)=k$.
Finally,
noticing that
integration over the variables $x_{i>l+1}$
 obviously gives 1,
in view of
the purely iterative definition of the process $Y_l$, which is independent of
the "future" $x_{i>l+1}$,
we have
\begin{equation}
\sum_{\pi \ {\rm generates}\
Y_l} p(\pi) = \sum_{\pi \ {\rm generates}\ Y_l}\frac{1}{(n+1)!},
\end{equation}
which is precisely the probability of the trajectory
$X_l^{(n)}$ with $l\le n$.

Therefore,
$Y_l$ and $X_l^{(n)}$
are identically distributed.
We emphasize that the distribution of $Y_l$, and therefore
 the distribution of $X_{l}^{(n)}$, do {\it not} depend
on $n$ for $l < n$, which is a rather counter-intuitive
result for the process
$X_{l}^{(n)}$. Consequently,  the
probability ${\cal P}(X_l^{(n)} = X)$ that at any intermediate step $l$, $l = 1,2,3, \ldots , n-1$,
the trajectory
$X_l^{(n)}$ appears at the site $X$ obeys
\begin{equation}
\label{v}
{\cal P}(X_l^{(n)} = X) = {\cal P}_l(X),
\end{equation}
where ${\cal P}_l(X)$ is defined by Eq.(\ref{eu}).

We note here parenthetically
that the result in Eq.(\ref{v})
allows us  to
study the
properties of the random process $X_l^{(n)}$
which
depend on the
intermediate states,  such as, e.g.,  the span, the maximal excursions within time $n$,
the time spent on positive
sites and so on.
In particular, it might be
instructive to compare the growth of the average length $L_n$ of
the longest subsequence of (not necessarily consecutive) rises,
i.e. of the so called "longest increasing subsequence", and the
growth of the average maximal positive excursion $< {\rm max}\{X_{n}\} >$
of the PGRW, which is also supported by rises in random
permutations. Since we realized that the PGRW converges in
distribution to standard P\'olya walk with diffusion coefficient
$D =1/6$, we may estimate $< {\rm max}\{X_{n}\} >$ as \cite{weiss}:
\begin{equation}
< {\rm max}\{X_{n}\} > \sim \left(\frac{4 D}{\pi}\right)^{1/2} \cdot \sqrt{n} =
\left(\frac{2}{3\pi}\right)^{1/2} \cdot \sqrt{n} \approx 0.461 \cdot
\sqrt{n}
\end{equation}
Hence, the growth of the maximal positive displacement of the
PGRW proceeds at a slower
rate  than that of $L_n$, $L_n \sim 2 \sqrt{n}$, due to a numerical
factor, which is more than four times smaller.

\subsection{Probability measure of different trajectories.}

The equivalence of the processes $Y_l$ and $X_l^{(n)}$
 allows us to determine
explicitly the
probability of any given trajectory.
We note that in the permutation language, this problem amounts to
the calculation of the number of permutations with a
\textit{given} rise-and-descent sequence and has been already
solved in terms of an elaborated combinatorial approach in
Refs.\cite{macmahon,carlitz,niven}. Here we propose a novel
solution of this problem.

Let $X_l^{(n)}$ be a given trajectory
generated by a random uniform permutation $\pi$ of $[n+1]$.
This trajectory, according to the definition of the PGRW,
can be uniquely defined in terms of the sequence
of rises ($\uparrow$) and descents ($\downarrow$) characterizing
the permutation $\pi$. Now, let $\hat{I}_\uparrow$ and $\hat{I}_\downarrow$
denote
the integral operators of the form
\begin{equation}
\hat{I}_\uparrow = \int_x^1 dx  \cdot \ {\rm and}\ \hat{I}_\downarrow = \int_0^x dx  \, \cdot
\end{equation}
Further on, to each
$l$-step trajectory $X_l^{(n)}$, we associate
a polynomial $Q_{X_l^{(n)}}(x) \in\R[x]$ of degree $l$, defined as
\be
\label{Q}
Q_{X_l^{(n)}}(x) =\prod_{i=1}^{l} \hat{I}_i \cdot 1 = \hat{I}_1 \hat{I}_2 \hat{I}_3 \cdots \hat{I}_l \cdot 1,
\end{equation}
where $\hat{I}_i$ assumes either of two values - $\hat{I}_\uparrow$ and $\hat{I}_\downarrow$,
prescribed by the
direction of the "arrow" at the $i$-th step in
the corresponding sequence. In particular,
the polynomial $Q_{X_l^{(n)}}(x)$ corresponding to the $l=5$-step trajectory
$X_l^{(n)} = \{\uparrow,\uparrow,\downarrow,\uparrow,\uparrow\}$ will be
\begin{eqnarray}
Q_{X_l^{(n)}}(x) &=& \hat{I}_\uparrow \hat{I}_\uparrow \hat{I}_\downarrow \hat{I}_\uparrow \hat{I}_\uparrow \cdot 1 = \nonumber\\
&=& \int_{x}^1 dx_1 \int_{x_1}^1 dx_2  \int_0^{x_2} dx_3     \int_{x_3}^1 dx_4 \int_{x_4}^1 dx_5 \cdot 1 = \nonumber\\
&=& \frac{3}{40} - \frac{x}{8} + \frac{x^3}{12} - \frac{x^4}{24} + \frac{x^5}{120}.
\end{eqnarray}
Then, the desired probability measure of a given trajectory $X_l^{(n)}$ is given by
\begin{equation}
\label{measure}
p(X_l^{(n)})=\int_0^1 Q_{X_l^{(n)}}(x) dx = \int_0^1 dx \prod_{i=1}^{l} \hat{I}_i \cdot 1
\end{equation}
Note that this probability measure is not homogeneous, contrary to the measure of the standard P\'olya walk.

The result in Eq.(\ref{measure})
may be compared with the analogous expression for the number of
permutations with a given rise-and-descent sequence
found by Niven \cite{niven}.
Following Niven, consider a fixed \text{up-and-down} arrow sequence
of length $l$ and denote by $k_1, k_2, \ldots, k_r$ the positions of downarrows ($r$ is the total number of downarrows) along the sequence.
Suppose next that this  \text{up-and-down} sequence corresponds to some random permutation $\pi$ of $[n+1]$ such that,
according to conventional notation,
an $up$-arrow represents a rise, while a $down$-arrow is a descent. A question now is
to calculate the \textit{number}
$N(X_l^{(n)})$
of permutations generating a given  \text{up-and-down} sequence  (or, in our language, a given trajectory $X_l^{(n)}$). Elaborated combinatorial
arguments show that $N(X_l^{(n)})$ equals the determinant of a matrix $A$ of order $r+1$ whose elements $\alpha_{i,j}$
(where $i$ stands for the row, while $j$ - for the column) are binomial coefficients $k_i \choose k_{j-1}$, where
$k_0=0,\ k_{r+1}=l+1$, and
${m \choose n}=0$ if $n>m$
\cite{niven}. Consequently, an alternative expression for the probability
measure
$p(X_l^{(n)})$ may be written down as
\begin{equation}
\label{det}
p(X_l^{(n)})= \frac{1}{(l + 1)!} {\rm det}
\begin{pmatrix}
\displaystyle
1 & 1& \displaystyle k_1 \choose k_2 & \displaystyle k_1 \choose k_3 & \ldots &  \displaystyle k_1 \choose k_r \\
1 & \displaystyle k_2 \choose k_1 & 1 & \displaystyle k_2 \choose k_3 & \ldots &  \displaystyle k_2 \choose k_r\\
1 & \displaystyle k_3 \choose k_1 & \displaystyle k_3 \choose k_2 & 1 & \dots & \displaystyle k_3 \choose k_r\\
&&&\ldots &\\
1 &  \displaystyle k_{r+1} \choose k_1 & \displaystyle k_{r+1} \choose k_2 & \displaystyle k_{r+1} \choose k_3 & \dots & \displaystyle k_{r+1} \choose k_r
\end{pmatrix}.
\end{equation}
One may readily verify that both Eq.(\ref{measure}) and Eq.(\ref{det}) reproduce our
 earlier results in Eqs.(\ref{p0}) to (\ref{p3}) determining the probabilities of different four step
trajectories.

Finally, similarly to our Eqs.(\ref{2}) and (\ref{4}), we may define a general, $k$-point
correlation function ${\cal C}^{(k)}_{j_1,..,j_k}$ of the \text{rise-and-descent} sequence:
\begin{equation}
{\cal C}^{(k)}_{j_1,..,j_k} = \Big \langle s_{j_1} s_{j_2}
s_{j_3} \cdots s_{j_k}\Big \rangle,
\end{equation}
where the "spin" variable $s_k$ has been defined in Eq.(\ref{spin}).
In terms of the auxiliary process $Y_l$,
${\cal
C}^{(k)}_{j_1,..,j_k}$
can be rewritten formally as
\be\label{correlation}
\displaystyle {\cal C}^{(k)}_{j_1,..,j_k}=\left<\prod_{i=1}^k
\theta(x_{j_i+1}-x_{j_i})\right>_{\{x_k\}}=\int_{0}^1dx_{j_1} \ldots \int_{0}^{1}dx_{j_k+1}\prod_{i=1}^k
\theta(x_{j_i+1}-x_{j_i}), \fin
where the brackets with the subscript $\{x_k\}$ denote
averaging with respect to the distribution of the ensemble of
variables $x_k$.

We note that the purely nearest-neighbors nature of the $k$-point
correlations, (which we have taken for granted in Section 4),
indeed becomes quite transparent for the process $Y_l$, such that
${\cal C}_{j_1,..,j_k}^{(k)}$  factorizes automatically into a
product of the corresponding correlation functions of the
consecutive subsequences, in which all $j_k$ differ by unity, as
soon as any of the distances $j_{k+1} - j_k $ exceeds unity. On
the other hand, the correlation function ${\cal C}^{(k)} = {\cal
C}^{(k)}_{j,j+1,..,j+k}$ of a consecutive sequence of arbitrary
order $k$  can be obtained recursively using
Eq.(\ref{correlation}). To do this, we first note that ${\cal
C}^{(k)}$ can be represented as \be {\cal C}^{(k)} = \int_0^{1} dx
f_k(x),
\end{equation}
where $f_k(x)$ are polynomials of order $k$, which obey the recursion
\begin{eqnarray}
f_k(x) = - \int^x_0 dx f_{k-1}(x) + \int^{1}_x dx f_{k-1}(x), \;\;\; f_0(x) = 1.
\end{eqnarray}
From this recursion, one finds immediately that ${\cal C}^{(k)}$ are defined through:
\begin{eqnarray}
{\cal C}^{(k)} = \sum_{p=0}^{k-1} \frac{(-1)^k 2^k}{(k+1)!} {\cal C}^{(k-1-p)}, \;\;\; {\cal C}^{(0)} = 1,
\end{eqnarray}
such that the generating function of ${\cal C}^{(k)}$ obeys:
\be
\label{jpt}
{\cal C}^{(k)}(z) \equiv \sum_{k=0}^{\infty} {\cal C}^{(k)} z^k = \frac{\tanh(z)}{z}.
\end{equation}
Consequently, ${\cal C}^{(k)}$ are related to the tangent numbers \cite{foata},
and are given explicitly by
\begin{equation}
\label{dddf}
{\cal C}^{(k)} = \frac{(-1)^k 2^{k+2} \left(2^{k+2} - 1\right)}{(k + 2)!} B_{k+2},
\end{equation}
where $B_k$ are the Bernoulli numbers. Note that since
the Bernoulli numbers equal zero
for $k$ odd, we find that
correlation functions of odd order vanish, i.e. ${\cal
C}^{(2 k + 1)} \equiv 0$.

Finally, using the asymptotic expansions for the factorial and the Bernoulli numbers,
we get that for $k \gg 1$, the $k$-point correlation functions ${\cal C}^{(k)}$ ($k$ is even)
decay as
\be
{\cal C}^{(k)} \sim 2 (-1)^{k/2} \left(\frac{2}{\pi}\right)^{k+2},
\end{equation}
i.e. the $k$-point correlations fall off
exponentially with $k$ with a characteristic length $= \ln^{-1}(\pi/2)$.

\section{Distribution of the number of "turns" of the PGRW trajectories.}

In this section we study an important measure of how scrambled the
PGRW trajectories are. This measure is
the number ${\cal N}$ of "turns" of an $n$-step PGRW trajectory, i.e. the number of
times when the walker changes the direction of its motion. We focus here on the \textit{asymptotic}
form of the distribution
 ${\cal P}({\cal N},n)$ of the number of "turns" of an $n$-step PGRW trajectory.

In the permutation language, each turn to the left (right), when the walker making a jump to the
right (left) at time moment $j$
jumps to the left (right) at the next time moment $j+1$
corresponds evidently
to a peak $\uparrow \downarrow$ (a through $\downarrow \uparrow$) of a given permutation $\pi$, i.e. a sequence
$\pi_{j} < \pi_{j+1} > \pi_{j+2}$ ($\pi_{j} > \pi_{j+1} < \pi_{j+2}$).
Consequently, the distribution function
${\cal P}({\cal N},n)$
of the number of "turns"
of the PGRW trajectory, (i.e. the probability that an $n$-step
PGRW trajectory has exactly ${\cal N}$ turns),
is just the distribution
of the sum of peaks and throughs. The latter can be defined apparently
using
the so-called peak numbers
$P(n+1,m)$ of Stembridge
(i.e., the number of permutations of $[n+1]$ having $m$ peaks), which obey the
following three-term recurrence \cite{stembridge}:
\be
\label{recurrence}
P(n+1,m) = (2 m + 2) P(n,m) + (n + 1 - 2 m) P(n, m - 1).
\end{equation}
Below we will evaluate the asymptotic form of
such a distribution function
using a different type of arguments,
other than the Stembridge formula in Eq.(\ref{recurrence}),
expressing ${\cal P}({\cal N},n)$
through the correlation functions of the \text{rise-and-descent} sequences.

Let ${\cal N}_p$ and ${\cal N}_t$ denote the
number of peaks (i.e., the number of configurations $\pi_j < \pi_{j+1} > \pi_{j+2}$)
and throughs (i.e., the number of
configurations $\pi_j > \pi_{j+1} < \pi_{j+2}$) in a random permutation of $[n+1]$, respectively.
These realization-dependent numbers can be written down explicitly as
\be
{\cal N}_p = \frac{1}{4} \sum_{j = 1}^{n-1} \left(1 + s_j\right) \left(1 - s_{j+1}\right),
\end{equation}
and
\be
{\cal N}_t = \frac{1}{4} \sum_{j = 1}^{n-1} \left(1 - s_j\right) \left(1 + s_{j+1}\right),
\end{equation}
where the "spin" variable $s_j$ has been defined in Eq.(\ref{spin}).
The total number of turns ${\cal N}$ of the RW trajectory,
generated by a given permutation, is then given by
\be
{\cal N} = {\cal N}_p + {\cal N}_t = \frac{1}{2} \sum_{j = 1}^{n-1} \left(1 - s_j  s_{j+1}\right)
\end{equation}
We seek now the characteristic function ${\cal Z}_n(k)$ of ${\cal N}$, defined as
\begin{equation}
\label{partition}
{\cal Z}_n(k) = \Big\langle \exp\left(i k {\cal N}\right) \Big \rangle = \exp\left( \frac{i k (n - 1)}{2} \right)
\Big\langle \exp\left(- \frac{i k}{2} \sum_{j=1}^{n-1} s_j s_{j+1}\right) \Big \rangle,
\end{equation}
where the angle brackets denote averaging with respect to random
permutations of $[n+1]$. Note that ${\cal Z}_n(k)$ in
Eq.(\ref{partition}) can be thought of as a partition function of
a somewhat \text{exotic} one-dimensional Ising-type model in which
the "spin" variables $s_{j}$ of different sites are functionals of
consecutive numbers in random permutations and thus possess rather
specific correlations.

Now, since $s_j s_{j+1} = \pm 1$, the characteristic function ${\cal Z}_n(k)$ in Eq.(\ref{partition})
can be written down as
\be
\label{part2}
{\cal Z}_n(k) = \left(\frac{1+e^{i k}}{2}\right)^{n-1} \Big \langle \prod_{j=1}^{n - 1} \left( 1 - t s_j s_{j+1}\right)\Big \rangle,
\end{equation}
where $t =  \tanh(i k/2)$. Averaging the product in Eq.(\ref{part2}) and
taking into account
that
\begin{itemize}
\item $s_{j}^2 \equiv 1$,
\item $k$-point correlations with $k$ - odd vanish,
\item the $k$-point correlations factorize into a product of
the correlation functions of corresponding subsequences, as soon as
the distance between any two points exceeds unity,
\end{itemize}
we find
that
${\cal Z}_n(k)$ are polynomial functions of $t$
of order $[n/2]$, where $[x]$ is the floor function,
of the following form
\be
\label{part3}
{\cal Z}_n(k) = \left(\frac{1+e^{i k}}{2}\right)^{n-1} \sum_{l=0}^{[n/2]} (-1)^l W_{l,n} \, t^l,
\end{equation}
in which expression $W_{l,n}$
denote the weights
of corresponding configurations. These weights
$W_{l,n}$ can be obtained via straightforward
combinatorial calculations and are given explicitly by:
\be
\label{weights}
W_{l,n} = \displaystyle \sum  { n - 2l + 1 \choose \sum_{j=1}^l m_j } \frac{\left(\sum_{j=1}^l m_j\right)!}{m_1! m_2! m_3! \ldots m_l!} \prod_{j = 1}^{l} \left( {\cal C}^{(2 j)} \right)^{m_j},
\end{equation}
for $l > 0$,
where the summation extends over all positive
integers $m_{j}$ obeying the equation: $m_1 + 2 m_2 + 3 m_3 + \ldots + l m_l = l$,
and
\be
W_{l=0,n} = 1.
\end{equation}
Note that in Eq.(\ref{weights}) the symbol ${ a \choose b}$,
in a standard fashion, denotes the binomial coefficient for $a \geq b$ and equals zero otherwise.

Now, it is expedient to introduce the generating function of the form:
\be
{\cal Z}(k,z) = \sum_{n = 2}^{\infty} {\cal Z}_n(k) \; z^n.
\end{equation}
Note that summation over $n$ is performed for $n \geq 2$, since ${\cal N} \equiv 0$ for $n = 1$ and $n = 0$.
Multiplying both sides of Eq.(\ref{part3}) by $z^n$ and performing summation, we get, (relegating the intermediate steps to Appendix B),
\begin{eqnarray}
\label{corrr}
{\cal Z}(k,z) &=& - \disp \frac{4}{\left(1 + e^{i k}\right)^2 z} - \frac{2}{1 + e^{i k}} - z + \nonumber\\
&+&\frac{4}{\left(1 + e^{i k}\right)^2 z} \left[1 - \frac{\left(1 + e^{i k}\right) z}{2} \sum_{j = 0}^{\infty}{\cal C}^{(2 j)} \left( \frac{\left(1 - e^{2 i k}\right) z^2}{4}\right)^{j}\right]^{-1}
\end{eqnarray}
Therefore, Eq.(\ref{corrr}) relates the characteristic function of the sum of peaks and throughs
in random permutations, (or, equivalently, of the number of turns of the PGRW trajectories), to the $k$-point correlations in the \text{rise-and-descent} sequences, Eq.(\ref{dddf}). Using next the expression in Eq.(\ref{jpt}), we find, eventually,
the following explicit result for the characteristic function ${\cal Z}(k,z)$:
\be
\label{final}
{\cal Z}(k,z) = \disp \frac{4}{\left(1 + e^{i k}\right)^2 z} \left[\left(\frac{1-e^{i k}}{1 + e^{i k}}\right)^{1/2} \coth\left(\left(1 - e^{2 i k}\right)^{1/2} \frac{z}{2}\right)-1\right]^{-1} - \frac{2}{1 + e^{i k}} - z.
\fin
Expanding next ${\cal Z}(k,z)$ in Taylor series in
powers of variable $k$, we get
\be
{\cal Z}(k,z) = \frac{z^2}{1 - z} + \frac{2 z^2 i k}{3 \left(1 - z\right)^2}
- \frac{z^2 \left(60 + 15 z + 6 z^2 - z^3\right)}{90 \left(1 - z\right)^3} \frac{k^2}{2} + {\mathcal O}\Big(k^3\Big),
\fin
which implies, in particular, that the average and averaged squared number of
turns of the PGRW trajectory are given by
\be
\Big< {\cal N} \Big> = \frac{2 (n - 1)}{3}
\fin
and
\be
\Big< {\cal N}^2 \Big> = \frac{\left(5 n^2 - 7 n + 2\right)}{12} \theta(n - 1) + \frac{\left(n-3\right)}{15} \theta(n - 3) + \frac{\left(n^2 - 7 n + 12\right)}{36} \theta(n - 4).
\fin
Consider finally the form of ${\cal Z}(k,z)$ in the limit of small $k$ and $z \to 1^{-}$. In this limit, we find from Eq.(\ref{final}) that in
the leading order ${\cal Z}(k,z)$ obeys:
\be
\label{invrt}
{\cal Z}(k,z) \sim \disp \frac{1}{1 - z - \disp \frac{2 i}{3} k + \disp \frac{4}{45} k^2}
\fin
Inverting Eq.(\ref{invrt}) with respect
to $k$ and $z$, we thus obtain that in the
asymptotic limit $n \to \infty$, the distribution function
${\cal P}({\cal N},n)$
of the number of "turns"
of the PGRW trajectory converges to a normal distribution of the form:
\be
{\cal P}({\cal N},n) \sim \frac{3}{4} \Big(\frac{5}{\pi n}\Big)^{1/2} \exp\left(- \frac{45 \Big({\cal N} - \disp \frac{2}{3} n\Big)^2}{16 n}\right).
\fin
This is not, of course, a counter-intuitive result in view of the polynomial
three-term recursion in Eq.(\ref{recurrence}) obeyed by (closely related to ${\cal P}({\cal N},n)$)
 peak numbers
$P(n+1,m)$ \cite{stembridge}.

\section{Diffusion limit}

Consider finally a continuous space and time version of
the PGRW in the diffusion limit.
To do this, it is expedient to define first
some sort of an "evolution" equation
for ${\cal P}_l(Y)$.

We proceed as follows. Define first
the polynomial: \be
\displaystyle V^{(l)}(x,Y)=\sum_{Y_l = Y} Q_{Y_l}(x),
\end{equation}
where the polynomial $Q_{Y_l}(x)$ has been determined in Eq.(\ref{Q}) and
the sum extends over all $l$-steps trajectories $Y_l$ starting at zero and
ending at the fixed point $Y$. Note next that one has
\be
\displaystyle {\cal
P}_l(Y)=\left<V^{(l)}(x,Y)\right>_{\{x_l\}}
\end{equation}
Now, for the polynomials $V^{(l)}(x,Y)$ one obtains, by counting all possible trajectories
$Y_l$ starting from zero and ending at the fixed point $Y$, the following "evolution" equation:
\be
\label{master}
V^{(l+1)}(x,Y)=\hat{I}_\uparrow \cdot V^{(l)}(x,Y-1)+\hat{I}_\downarrow \cdot V^{(l)}(x,Y+1).
\end{equation}
Taking next advantage of the established
equivalence between the processes $Y_l$ and
$X_l^{(n)}$, we can rewrite the last equation, upon averaging
it over the distribution of variables $\{x_l\}$, as
\be
\label{master2}
{\cal P}_{l+1}(Y)=\frac{(l-Y+4)}{2(l+1)}{\cal
P}_{l}(Y-1)+\frac{(l+Y)}{2(l+1)}{\cal P}_{l}(Y+1),
\end{equation}
which represents
the desired evolution equation for ${\cal P}_{l}(Y)$ in discrete space and time.

We hasten to remark that Eq.(\ref{master2}) can be thought of as
a direct consequence of the celebrated relation between
the Eulerian numbers:
\be \label{eulerian} \disp
\euler{n+1}{l}=l
\left \langle
\begin{matrix}
n \\ \displaystyle l
\end{matrix}
\right \rangle
+(n+2-l)\euler{n}{l-1}.
\end{equation}
We emphasize, however,
that despite the fact such a relation is evident for the distribution
${\cal P}_l(X_l^{(l)}=X)$ of the end-point of the PGRW trajectories, Eq.(\ref{eu}),
its generalization to the
distribution of the \text{intermediate} steps ${\cal P}_l(X_l^{(n)}=X)$
as well as to
the distribution of the process $Y_l$,  ${\cal P}_l(Y=X)$, is
a completely unevident \textit{a priori} result,
which enlightens the process under
consideration.

Given the evolution Eq.(\ref{master2}), we turn next to
the diffusion limit. Introducing space $y = a Y$ and time $t =
\tau l$ variables, where $a $ and $\tau$ define characteristic
space and time scales, we turn to the limit $a, \tau \to 0$,
supposing that the ratio $a^2/\tau$ remains fixed and determines
the diffusion coefficient $D_0 = a^2/ 2 \tau$. In this limit,
Eq.(\ref{master2}) becomes \be\label{fokker}
\frac{\partial}{\partial t}{\cal P}(y,t)=\frac{\partial}{\partial
y}\left(\frac{y}{t}{\cal
P}(y,t)\right)+D_0\frac{\partial^2}{\partial y^2}{\cal P}(y,t)
\fin Note that the resulting continuous space and time equation is
of the Fokker-Planck type; it has a constant diffusion coefficient
and a negative drift term which, similarly to the
Ornstein-Uhlenbeck process, grows linearly with $y$, but the
amplitude of the drift decays in proportion to the first inverse
power of time, which signifies that the process $Y_l$ eventually
delocalizes. Note also that the form of Eq.(\ref{fokker}) ensures
the conservation of the total probability.

The Green's function solution of Eq.(\ref{fokker}),  remarkably, is a normal distribution
\be
{\cal P}(y,t)=\sqrt{\frac{3}{4\pi D_0 t}} \exp\left(- \frac{3 y^2}{4 D_0t}\right) \fin
which is consistent with the large-$n$
limit of the discrete
process derived in Eq.(\ref{normal}).

Now, the Langevin equation corresponding to Eq.(\ref{fokker})
reads\be\label{langevin} \frac{dy}{dt}=-\frac{y}{t}+\zeta(t) \fin
with $\left<\zeta(t)\zeta(t')\right>=2D_0\delta(t-t')$. Its
solution can be readily obtained and has the following form:
\be\label{solange} y(t)=\frac{1}{t}\int_0^t t'\zeta(t')dt' \fin
Note that Eqs.(\ref{fokker}) or (\ref{langevin}) model dynamics of
an overdamped particle in a one-dimensional continuum subject to a
force $-y/t$ and a white noise. Note also that Eq.(\ref{solange})
implies (see Ref.\cite{yor}) that $y(t)$, being a linear
functional of a white noise, can be represented as a rescaled
Brownian motion of the form: \be y(t)=\int_0^t\zeta'(t')dt' \fin
where $\left<\zeta'(t)\zeta'(t')\right>=2D_0\delta(t-t')/3$.
Consequently, $y(t)$ can be also thought of as the solution of the
customary Langevin equation with zero external force, \be
\label{lang} \frac{dy}{dt}=\zeta'(t). \fin We finally remark that
Eq.(\ref{lang}) could have been derived directly from the
definition of the discrete process $Y_l$, by taking the proper
continuum limit of Eq.(\ref{Y}). The correlator
$\left<\zeta'(t)\zeta'(t')\right>$ must be taken in this case as
the limit of the continuation ${\tilde {\cal C}}^{(2)}(m)$ of
${\cal C}^{(2)}(m)$ to $m<0$: \be
\left<\zeta'(t)\zeta'(t')\right>=\lim_{a,\tau\to0}{\tilde{\cal
C}}^{(2)}(m)=\lim_{a,\tau\to0}\left(\delta_{m,0}-\frac{1}{3}(\delta_{m,-1}+\delta_{m,1})\right)=2D_0\delta(t-t')/3
\fin The Gaussian nature of the process $y(t)$ is transparent
within this formalism.

\section{Conclusions}

In conclusion, we have
studied here a simple model of a non-Markovian random walk,
evolving in discrete
time on a one-dimensional lattice of integers, in which the moves
of the walker to the right or to the left are prescribed by the
\text{rise-and-descent} sequence characterizing each random
permutation $\pi$ of $[n+1]=\{1,2,3, \ldots, n+1\}$.
We have determined exactly the probability ${\cal P}_n(X)$ of finding the
end-point $X_n$ of the walker's trajectory  at site $X$.
Furthermore, we have shown that in the long-time limit ${\cal P}_n(X)$
converges to a normal distribution in which an effective diffusion
coefficient $D = 1/6$ is three times smaller that the diffusion
coefficient ($D = 1/2$) of the conventional one-dimensional P\'olya random walk.
This implies that correlations in the generator of the walk - random permutations,
which arise because only a finite amount of numbers is being shuffled
and neither of these number may be equal to each other,
are marginally important. Indeed,
we have shown that two- and -four-point
correlations in the sequence of rises depend only on the relative
distance between them
and
extend
 effectively
to nearest-neighbors only.
Next, at a closer look on the
intrinsic features of the PGRW trajectories, we have formulated an auxiliary
stochastic Markovian process. We have
shown that, despite the fact that this process is Markovian,
while the random walk generated by permutations
is not a Markovian process, the distribution of the
auxiliary stochastic process appears to be identic to the
distribution of the intermediate states of the walker's trajectories.
This enabled us to obtain the probability measure of different
excursions of the permutations-generated random walk,
determine general $k$-point correlation functions and to evaluate
the asymptotic form of the probability distribution of the number of turns in an $n$-step PGRW trajectory.
Finally, we have discussed, in the diffusion
limit, the continuous space and time version of such a walk.

\section{Acknowledgments}

The authors gratefully acknowledge helpful discussions with M.N.
Popescu, S. Dietrich and R. Metzler. We also wish to thank
S.Nechaev for many valuable comments and
for pointing us on the paper by Stembridge \cite{stembridge}.
R.V. thanks
Max-Planck Institute Stuttgart for warm hospitality and G.O.
thanks the Alexander von
Humboldt Foundation for the financial support via the Bessel
Research Award.

\newpage

\section{Appendix A}.

Here we briefly outline
the derivation of the integral representation of ${\cal P}_n(X)$
in Eq.(\ref{br1}). Using the following expansion of the polylogarithm function \cite{abr}:
\begin{equation}
{\rm Li}_{-n-1}\Big(\exp(- 2 i k)\Big) = \frac{(n + 1)!}{\displaystyle (2 \pi i)^{n+2}} \sum_{q = -\infty}^{\infty}
\frac{1}{\displaystyle \left(q +
\frac{k}{\pi}\right)^{n+2}}
\end{equation}
we have that $\tilde{{\cal P}}_n(k)$ in Eq.(\ref{dd}) can be cast
into the form
\begin{equation}
\tilde{{\cal P}}_n(k) = \frac{\sin^{n+2}(k)}{\pi^{n+2}}
\sum_{q=-\infty}^{\infty} \frac{1}{\displaystyle \left(q +
\frac{k}{\pi}\right)^{n+2}}
\end{equation}
Noticing next that
\begin{eqnarray}
\sum_{q=-\infty}^{\infty} \frac{1}{\displaystyle \left(q +
\frac{k}{\pi}\right)^{n+2}} &=& \frac{1}{(n+1)!} \left[
\Psi_{n+1}\left(1-\frac{k}{\pi}\right) +
(-1)^{n} \Psi_{n+1}\left(\frac{k}{\pi}\right)\right] \nonumber\\
&=& \frac{(-1)^{n+1} \pi^{n+2}}{(n+1)!} \frac{d^{n+1}}{d k^{n+1}}
\cot(k),
\end{eqnarray}
where $\Psi_n(x)$ denotes the polygamma function \cite{abr}, we arrive eventually at the
representation in Eq.(\ref{br1}).

\newpage

\section{Appendix B}

Here we present the details of the derivation of the generating function in Eq.(\ref{corrr}).
Multiplying both sides of Eq.(\ref{part3}) by $z^n$ and performing summation, we get
\begin{eqnarray}
{\cal Z}(k,z) &=& \disp \frac{2}{1+e^{i k}} \sum_{l = 0}^{\infty} \left(\frac{1- e^{i k}}{1+e^{i k}}
\right)^l \sum_{n = 2 l}^{\infty} W_{l,n} \left(\frac{z \left(1 + e^{i k}\right)}{2}
\right)^n - \frac{2}{1+e^{i k}} - z = \nonumber\\
&=& \disp \frac{2}{1+e^{i k}} \sum_{l = 0}^{\infty} \left(\frac{1- e^{i k}}{1+e^{i k}}
\right)^l \left(\sum   \frac{\left(\sum_{j=1}^l m_j\right)!}{m_1! m_2! m_3! \ldots m_l!} \prod_{j = 1}^{l} \left( {\cal C}^{(2 j)} \right)^{m_j}\right) \times \nonumber\\
&\times& \sum_{n=2l}^{\infty} { n - 2l + 1 \choose \sum_{j=1}^l m_j } \left(\frac{z \left(1 + e^{i k}\right)}{2}
\right)^n - \frac{2}{1+e^{i k}} - z =\nonumber\\
&=& \disp \frac{2}{1+e^{i k}} \sum_{l = 0}^{\infty} \left(\frac{1 - e^{2 i k}}{4} z^2\right)^l
\left(\sum   \frac{\left(\sum_{j=1}^l m_j\right)!}{m_1! m_2! m_3! \ldots m_l!} \prod_{j = 1}^{l} \left( {\cal C}^{(2 j)} \right)^{m_j}\right) \times \nonumber\\
&\times& \sum_{p = 0}^{\infty} {p+1 \choose \sum_{j=1}^l m_j} \left(\frac{z \left(1 + e^{i k}\right)}{2}
\right)^p - \frac{2}{1+e^{i k}} - z
\end{eqnarray}
Note now that ${p+1 \choose \sum_{j=1}^l m_{j}} \equiv 0$ for $p < \sum_{j=1}^l m_{j} - 1$. Hence, for $l > 0$, (when, evidently,  $\sum_{j=1}^l m_j \geq 1$), we have
\begin{eqnarray}
\label{lk}
\sum_{p = 0}^{\infty} {p+1 \choose \sum_{j=1}^l m_j} \left(\frac{z \left(1 + e^{i k}\right)}{2}
\right)^p &=&  \disp \sum_{p = \sum_{j=1}^l m_j - 1}^{\infty} {p+1 \choose \sum_{j=1}^l m_j} \left(\frac{z \left(1 + e^{i k}\right)}{2}
\right)^p = \nonumber\\
&=& \disp \left(\frac{z \left(1 + e^{i k}\right)}{2}
\right)^{\sum_{j = 1}^l m_j - 1} \left(1 - \frac{z \left(1 + e^{i k}\right)}{2}\right)^{-1-\sum_{j=1}^l m_j}
\end{eqnarray}
On the other hand, for $l = 0$ (when $\sum_j m_j \equiv 0$),
\begin{eqnarray}
\label{lllk}
\sum_{p = 0}^{\infty} {p+1 \choose 0} \left(\frac{z \left(1 + e^{i k}\right)}{2}
\right)^p = \disp  \left(1 - \frac{z \left(1 + e^{i k}\right)}{2}\right)^{-1}.
\end{eqnarray}
Making use next of Eqs.(\ref{lk}) and (\ref{lllk}), as well as of the identity
\be
\displaystyle \sum^{\infty}_{l=1} \tau^l \sum \frac{\left(\sum_{j=1}^l m_j\right)!}{m_1! m_2! m_3! \ldots m_l!} \prod_{j = 1}^{l} x_j^{m_j} = \disp \frac{1}{1 - \sum_{j = 1}^{\infty} x_j \tau^j} - 1,
\end{equation}
we arrive eventually at the result in Eq.(\ref{corrr}).

\end{document}